# Cm$^2$-Scale Synthesis of MoTe$_2$ Thin Films with Large Grains and Layer Control

*David J. Hynek, Raivat M. Singhania, Shiyu Xu, Benjamin Davis, Lei Wang, Milad Yarali, Joshua V. Pondick, John M. Woods, Nicholas C. Strandwitz, and Judy J. Cha\**

D. J. Hynek, S. Xu, Dr. L. Wang, Dr. M. Yarali, J. V. Pondick, Dr. J. M. Woods, Prof. J. J. Cha*
Energy Sciences Institute, Department of Mechanical Engineering and Materials Science, Yale University, West Haven, CT 06516, United States
E-mail: judy.cha@yale.edu

R. M. Singhania, B. Davis, Prof. N. C. Strandwitz
Department of Materials Science and Engineering, Lehigh University, Bethlehem, PA 18015, United States

## Abstract

Owing to the small energy differences between its polymorphs, MoTe$_2$ can access a full spectrum of electronic states, from the 2H semiconducting state to the 1T′ semimetallic state, and from the T$_d$ Weyl semimetallic state to the superconducting state in the 1T′ and T$_d$ phase at low temperature. Thus, it is a model system for phase transformation studies as well as quantum phenomena such as the quantum spin Hall effect and topological superconductivity. Careful studies of MoTe$_2$ and its potential applications require large-area MoTe$_2$ thin films with high crystallinity and thickness control. Here, we present cm$^2$-scale synthesis of 2H-MoTe$_2$ thin films with layer control and large grains that span several microns. Layer control is achieved by controlling the initial thickness of the precursor MoO$_x$ thin films, which are deposited on sapphire substrates by atomic layer deposition and subsequently tellurized. Despite the van der Waals epitaxy, the precursor-substrate interface is found to critically determine the uniformity in thickness and grain size of the resulting MoTe$_2$ films: MoTe$_2$ grown on sapphire show uniform films while MoTe$_2$ grown on amorphous SiO$_2$ substrates form islands. This synthesis strategy decouples the layer control from the variabilities of growth conditions for robust growth results, and is applicable to grow other transition metal dichalcogenides with layer control.

Transition metal dichalcogenides (TMDs) have been extensively studied over the last decade for their layer-dependent physical and electrochemical properties that can be used for optoelectronic, flexible, and energy-harvest and storage devices.[1–4] In addition, TMDs provide a platform to explore a large electronic phase space as several structural phases can be accessed with each phase associated with a distinct electronic state.[5] Tellurides, such as MoTe$_2$ and WTe$_2$, have demonstrated almost the entire spectrum of electronic phases, including superconductivity,[6–8] quantum spin Hall state,[9,10] Weyl semimetallic state,[11,12] semimetallic state, and semiconducting state. MoTe$_2$ in particular has shown promise for phase-change memory applications due to the small energy difference between the 2H and 1T′ phase[13] with experimental demonstrations of the phase switching induced by strain,[14] gating,[15,16] and heating.[17] Despite the growing interests in the tellurides, large-scale synthesis of telluride thin films with precise layer control remains a challenge, although such synthesis has been demonstrated for sulfides and selenides.[18–21]

For transport studies, mechanical exfoliation can yield clean MoTe$_2$ flakes from bulk crystals.[3] However, mechanical exfoliation does not provide control over the thickness, size, or shape of the flakes and the yield is low. Recent synthesis efforts have shown MoTe$_2$ monolayer flakes that are a few microns large.[22,23] Large-area thin films of MoTe$_2$ have been demonstrated using molecular beam epitaxy,[24] chemical vapor deposition,[13,25–28] and solution phase synthesis,[29] albeit for a limited range of thicknesses. Robust growth results of MoTe$_2$ thin films from monolayer to *any* arbitrary thickness with large grains have yet to be demonstrated. In this work, we convert MoO$_x$ (2<x<3) thin films deposited on c-plane sapphire (0001) substrates by atomic layer deposition (ALD) to achieve cm$^2$-scale synthesis of MoTe$_2$ films with thickness down to a monolayer and large grains. The use of ALD precursor films enables thickness control and scalable growth, and decouples the thickness control from variations in growth conditions, leading to reproducible and robust growth results. The low reactivity of Te is overcome by the generation of the chemical intermediate hydrogen telluride, which readily coverts MoO$_x$ to MoTe$_2$. We observe a direct correlation between the number of ALD cycles of the MoO$_x$ films and the number of layers in the resulting MoTe$_2$ films, ranging from 8 layers down to a monolayer. Large grains of 2H-MoTe$_2$ are achieved based on nucleation-limited growth of the 2H phase out of an initial 1T′-MoTe$_2$ film with continual supply of tellurium, as previously reported.[13,30]

**Results and Discussion**

Amorphous MoO$_x$ thin films were deposited on sapphire substrates by ALD with cycle numbers ranging from 5 to 135 cycles (Methods). MoO$_x$ films were chosen as the precursor to MoTe$_2$ over Mo to minimize strain during conversion, as volume expansion from MoO$_3$ to MoTe$_2$ is ~ 8 times smaller than that from Mo to MoTe$_2$.[30,31] Low-energy ion scattering spectroscopy (LEIS) was performed on the MoO$_x$ films to determine the surface coverage of MoO$_x$ on sapphire. At 5 ALD cycles, MoO$_x$ mostly covers the substrate; at 14 cycles, the coverage is complete and no Al peaks from the sapphire substrate are observed in the LEIS data (Figure S1, Supporting Information).

The oxide films were placed in a two-zone tube furnace and heated to 580 °C while Te source powder was placed upstream and heated to 570 °C in the presence of $H_2$ gas at atmospheric pressure (**Figure 1a**). The vaporized Te reacts with $H_2$ to form $H_2Te$, which converts $MoO_x$ to $MoTe_2$ (Methods).[32] Figure 1b and 1c show the cross-section transmission electron microscope (TEM) images of a 14-cycle $MoO_x$ film and the converted bilayer $MoTe_2$ film on a sapphire substrate, prepared by focused ion beam milling (Methods). The synthesized $MoTe_2$ films are in the 2H phase, as characterized by plan-view TEM (Figure 1d-g). While the bright-field TEM image shows interesting contrast variations (Figure 1d), the selected-area electron diffraction from the same region shows presence of two grains (Figure 1d inset). Dark-field (DF) TEM images formed from the two diffraction spots (Figure 1e, f) reveal a clear boundary between the two grains. The contrast variations are thus most likely due to crumpling of the film caused by the transfer process and are not indicative of the film quality. Much weaker diffraction spots are observed in between the two sets of diffraction spots, suggesting potentially a third grain in the field of view of 4 $\mu m^2$ of $MoTe_2$. Figure S2 provides a complete DF-TEM mapping of the grain shown in Figure 1f, which is ~ 10 $\mu m$ wide. We note that for plan-view TEM, $MoTe_2$ films were grown on $SiO_2$ substrates for easy lift-off using hydrofluoric acid as etching of sapphire for lift-off was proven difficult. Figure 1h shows optical images of $MoO_x$ and $MoTe_2$ films of varying thicknesses on sapphire before and after tellurization, which demonstrates successful fabrication of $cm^2$-scale $MoTe_2$ films on sapphire.

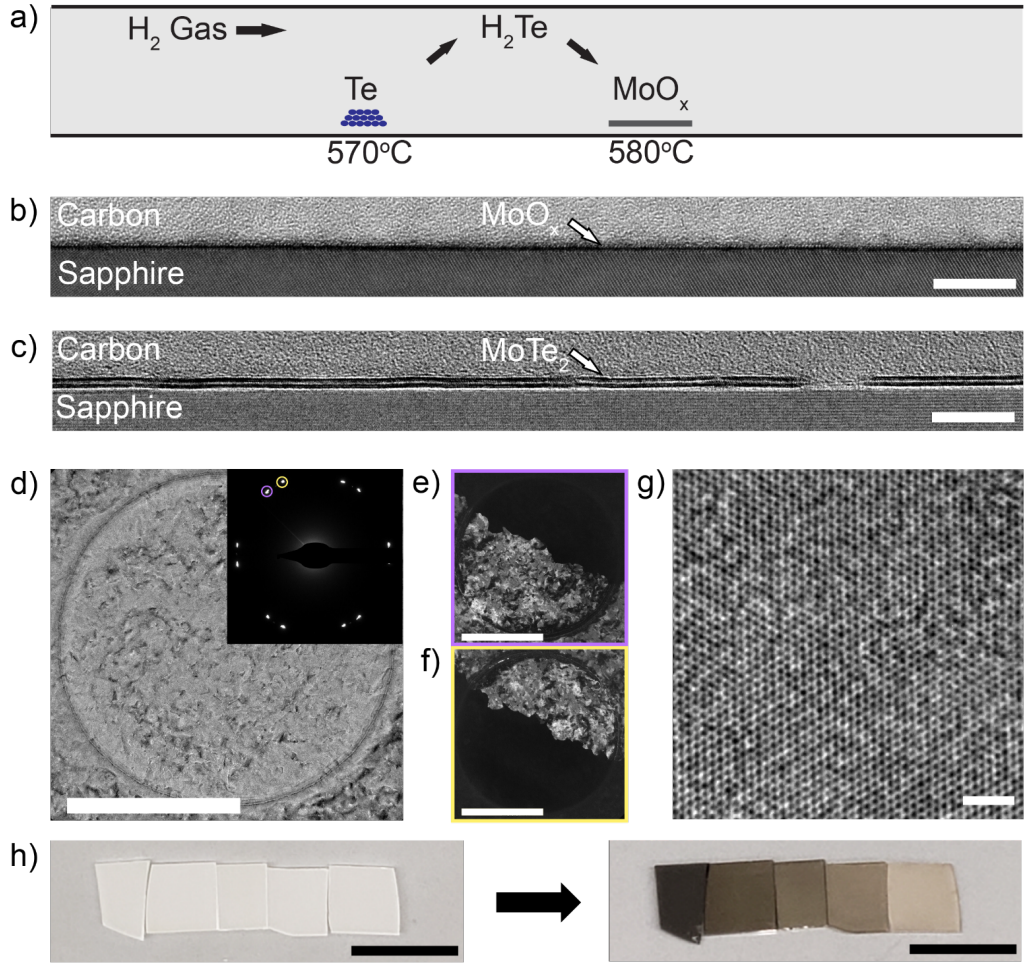

**Figure 1: Growth overview of MoTe$_2$ thin films. a)** Two-zone furnace growth schematic. **b, c)** TEM cross-section images of a 14-cycle MoO$_x$ film (**b**) and converted MoTe$_2$ film (**c**) on sapphire substrates. Scale bars, 10 nm. **d)** TEM image of a 2H MoTe$_2$ film grown on SiO$_2$, synthesized at 580 °C for 3 hours. Inset: The selected area electron diffraction pattern shows only two grains in field of view. Scale bar, 1 μm. **e, f)** Dark-field TEM images of the two grains present in (**d**), obtained from selecting the diffraction spot circled in purple (**e**) and yellow (**f**), respectively. Scale bar, 1 μm. **g)** High-resolution TEM image of the MoTe$_2$ film shows the expected hexagonal lattice of the 2H phase. Scale bar, 2 nm. **h)** Optical images of (left) MoO$_x$ thin films deposited on sapphire; from left to right corresponds to 135, 60, 27, 14, and 5 ALD cycles, and (right) the corresponding MoTe$_2$ films after tellurization. Scale bar, 1 cm.

**Figure 2** shows cross-section TEM images of MoTe$_2$ films of varying thickness: monolayer (Figure 2a), bilayer (Figure 2b), 2-3 layers (Figure 2c), 3-4 layers (Figure 2d), and 7-8 layers (Figure 2e). The converted MoTe$_2$ films are uniform in their thickness within a single layer step, demonstrating a high degree of uniformity across the substrate. Thus, the conversion from MoO$_x$ to MoTe$_2$ must be layer-by-layer and the number of layers of MoTe$_2$ can be controlled by the ALD cycle numbers of MoO$_x$. For the films shown in Figure 2, the corresponding ALD cycle numbers of the initial MoO$_x$ films are 5, 14, 27, 60, and 135 where a full surface coverage of the oxide film on sapphire occurs between 5 and 14 cycles (Figure S1, Supporting Information). There is a good correlation between ALD cycle number and MoTe$_2$ film thickness: the number of MoTe$_2$ layers increases monotonically with the number of ALD cycles (Figure S3, Supporting Information).

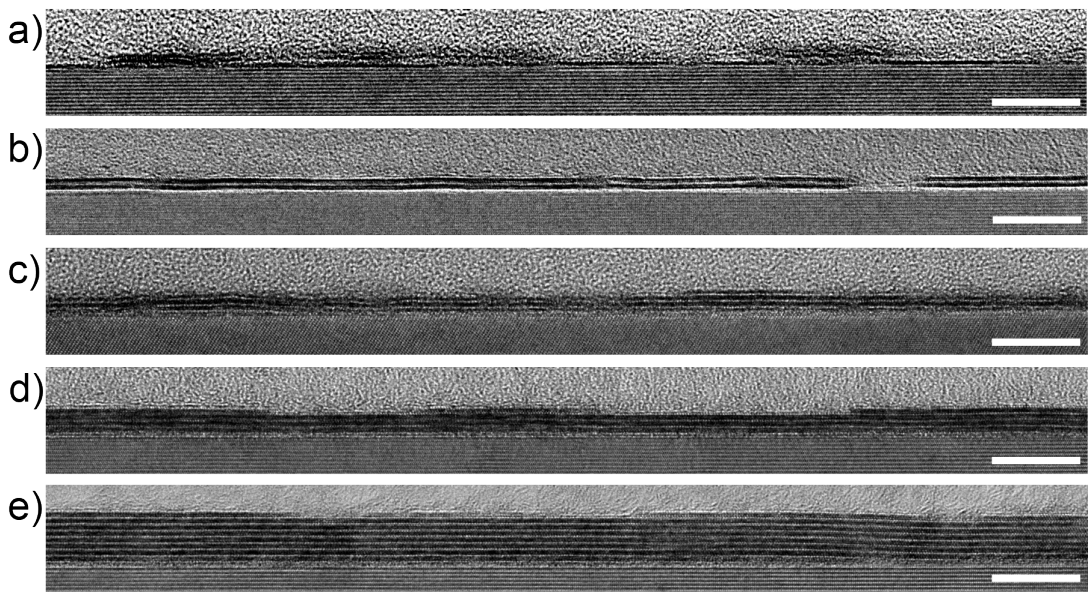

**Figure 2: Cross-sections of 2H MoTe$_2$ films on sapphire.** Cross-section TEM images of monolayer (**a**), bi-layer (**b**), 2-3 layer (**c**), 3-4 layer (**d**), and 7-8 layer (**e**) MoTe$_2$ thin films converted from 5, 14, 27, 60, and 135 cycle MoO$_x$ films deposited on sapphire, respectively. Scale bars, 10 nm.

The 2H phase of the MoTe$_2$ films was confirmed by Raman spectroscopy (**Figure 3a**). In agreement with prior Raman studies of exfoliated MoTe$_2$ flakes,[33–35] the intensities of the out-of-plane $A_{1g}$ (170 cm$^{-1}$) and bulk-inactive $B_{2g}^1$ (290 cm$^{-1}$) Raman modes are sensitive to layer number (Figure S4, Supporting Information), with the $A_{1g}$ peak becoming dominant over the in-plane $E_{2g}^1$ (234 cm$^{-1}$) peak in monolayer. The MoTe$_2$ film converted from the 5 cycle MoO$_x$ film shows a dominant $A_{1g}$ peak, verifying that the film is monolayer in agreement with the cross-section TEM result (Figure 2a). Because the $A_{1g}$ and $B_{2g}^1$ modes are sensitive to layer number, we examined the $A_{1g}$ and $B_{2g}^1$ modes at various locations of the films across a 5 – 10 mm region, which confirmed thickness uniformity at wafer scale (Figure S5, Supporting Information). Chemical analysis of the converted films using X-ray photoelectron spectroscopy shows a full conversion to MoTe$_2$ as evidenced by the peak shift of the Mo 3d peaks from MoO$_x$ to MoTe$_2$ (Figure 3b).[28] Energy dispersive x-ray spectroscopy shows that the stoichiometry of our MoTe$_2$ films is identical to that of a flake exfoliated from bulk MoTe$_2$ grown by chemical vapor transport (Figure 3c, Methods).

Surface uniformity of the films was further characterized by atomic force microscopy (AFM) (**Figure 4;** additional data in Figure S6, Supporting Information). For the monolayer MoTe$_2$ (Figure 4a), the AFM image shows patchy monolayer coverage across the substrate. This is attributed to insufficient precursor from the 5 ALD cycle MoO$_x$ film to yield a complete monolayer MoTe$_2$ film. The AFM observation agrees with the cross-section TEM image of the monolayer case (Figure 2a). The MoTe$_2$ converted from the 14-cycle MoO$_x$ shows a mostly uniform MoTe$_2$ film with pinholes of ~ 2 nm in depth (Figure 4b), which indicates that the film is bilayer, in agreement with the cross-section TEM result (Figure 2b). The presence of pinholes suggests that 14 ALD cycles of MoO$_x$ is close to the optimal cycle number for a complete bilayer film, but falls just short. For larger-cycle MoO$_x$ films (> 27 cycles) (Figure 4c, d), converted MoTe$_2$ films are uniform with no visible pinholes on the surface, and the height of the islands indicates one layer. These step heights are consistent with variations seen in cross-section TEM images (Figure 2c, d), and are uniformly present throughout the substrate (Figure S6, Supporting Information). Based on the AFM topographic images, 27 and 60 cycle MoO$_x$ films yielded ~ 40% areal coverage of the topmost layer in both cases, resulting in ~2.4 and 3.4-layer thick MoTe$_2$ films, respectively.

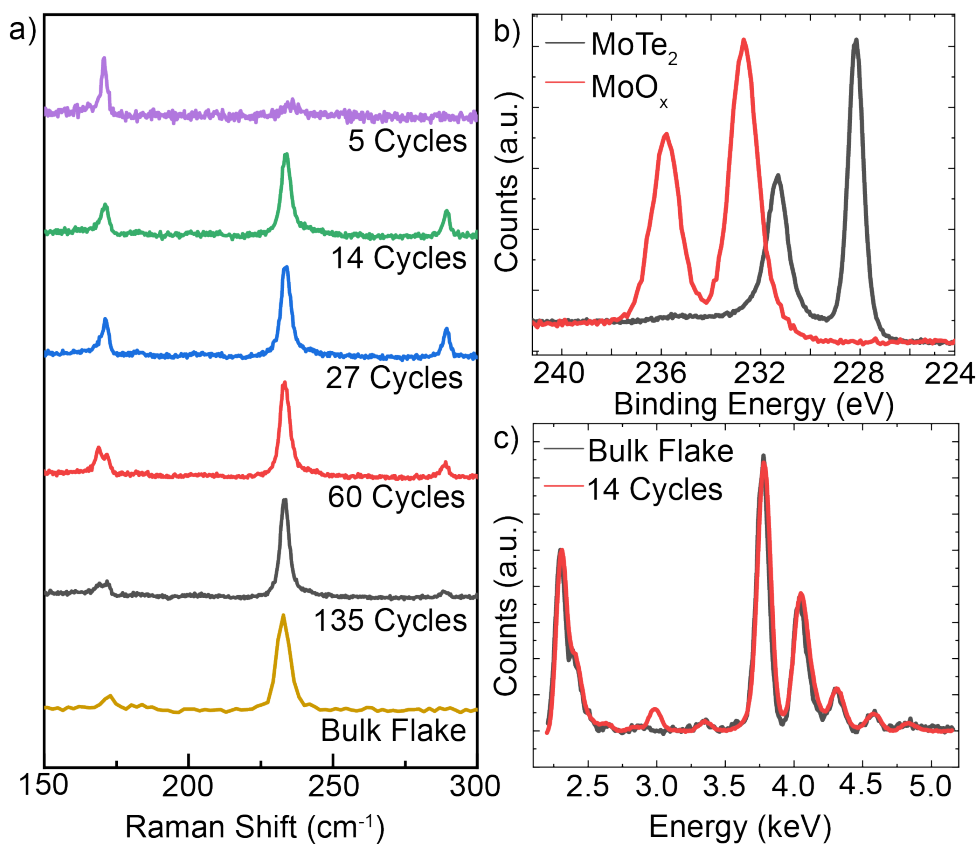

**Figure 3: Spectroscopic data from 2H MoTe₂ films. a)** Raman spectra of 2H MoTe₂ on sapphire using a 633 nm laser. Different cycle numbers correspond to different initial oxide thicknesses. The MoTe₂ film converted from the 5 cycle MoO$_x$ is monolayer based on the higher intensity of the $A_{1g}$ mode (170 cm$^{-1}$) compared to the $E_{2g}^1$ mode (234 cm$^{-1}$). **b)** X-ray photoelectron spectroscopy showing Mo *3d* peaks from MoO$_x$ films on sapphire before (red) and after (black) tellurization. The observed chemical shift indicates a full conversion from the oxide to the telluride. **c)** Normalized energy dispersive x-ray spectroscopy data of a MoTe₂ film converted from a 14 cycle MoO$_x$ thin film (red), overlaid with a reference spectrum of a bulk flake (black), showing that the stoichiometry for the converted film is identical to bulk crystal MoTe₂.

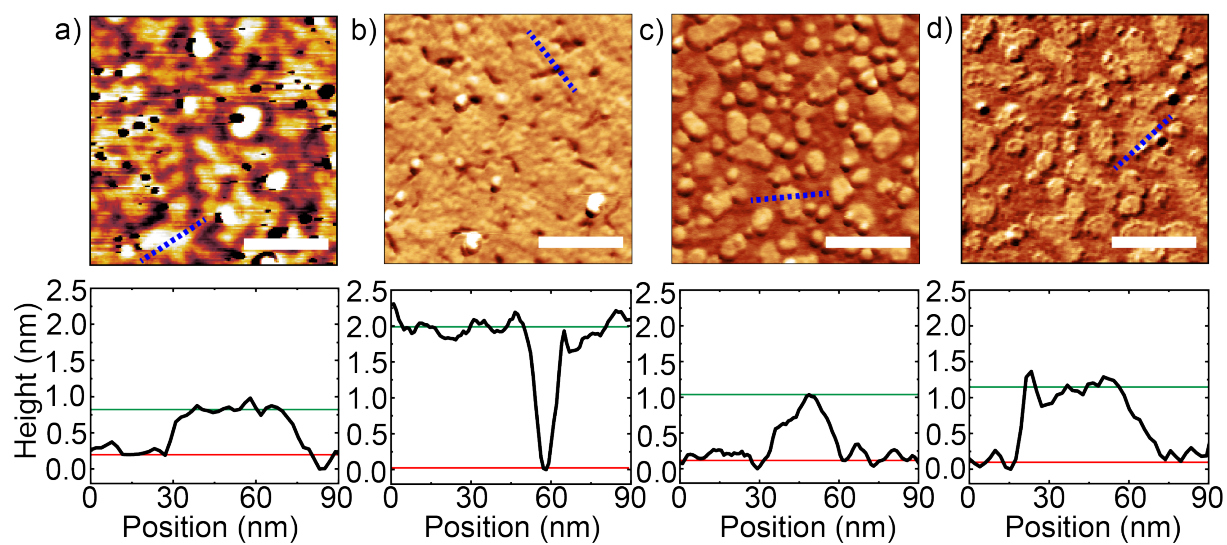

**Figure 4: Atomic force microscopy of MoTe$_2$ thin films.** AFM images and corresponding height profiles of MoTe$_2$ thin films converted from **a)** 5 cycle, **b)** 14 cycle, **c)** 27 cycle, and **d)** 60 cycle MoO$_x$ films. Scale bars, 100 nm. The height profiles were taken from the blue dotted lines marked in the AFM images. They show islands with ~ 1 nm step height, which corresponds to the thickness of one layer of MoTe$_2$. For the MoTe$_2$ shown in **b**, ~ 2 nm-deep pits are observed rather than islands, which correspond to pinholes in the bilayer MoTe$_2$.

Such analysis allows us to estimate the ideal ALD cycle number that will yield a MoTe$_2$ film with no variations in thickness across an entire substrate. For example, based on fits for low-cycle films, we estimate that 46 cycle MoO$_x$ would yield a complete tri-layer MoTe$_2$ film (Figure S3, Supporting Information).

The synthesized 2H-MoTe$_2$ films show large grains due to the nucleation-limited synthesis of the 2H phase.[13] In the early stage of the tellurization, MoTe$_2$ films are in the 1T′ phase. **Figure 5a** and 5b show TEM images of a MoTe$_2$ film that was tellurized for 1 hr instead of 4 hrs. A clear boundary is observed between 2H and 1T′ areas. The selected area diffraction pattern (Figure 5c) shows the presence of a single-crystalline 2H grain and poly-crystalline 1T′ regions with small nanoscale grains as evidenced by the ring diffraction patterns (high-resolution TEM image of 1T' phase in Figure S7, Supporting Information). Using Raman spectroscopy, we followed the conversion progression of the MoTe$_2$ films. We observe that MoTe$_2$ is initially in the 1T′ phase, then 2H grains nucleate out of the 1T′ phase and grow. This is tracked optically as the 2H phase appears lighter relative to the 1T′ phase. By changing the reaction temperature and thus controlling the nucleation and growth rate of 2H grains, the 2H grain size ranged from 20 um (Figure S2, Supporting Information) to 1 mm (Figure 5d). With longer reaction time, the 2H grains continue to grow and eventually merge, resulting in a complete 2H MoTe$_2$. A schematic of the phase conversion is shown in Figure 5g. The nucleation-limited synthesis enables synthesis of large-grain 2H-MoTe$_2$ thin films. Our results agree with previous studies that show the same synthesis pathway,[13] which suggests the initial 1T′ phase is due to tellurium deficiency at the beginning of tellurization when Te begins to replace oxygen in MoO$_x$. With sufficient time for tellurization, a continual supply of Te leads to the nucleation of the 2H phase.

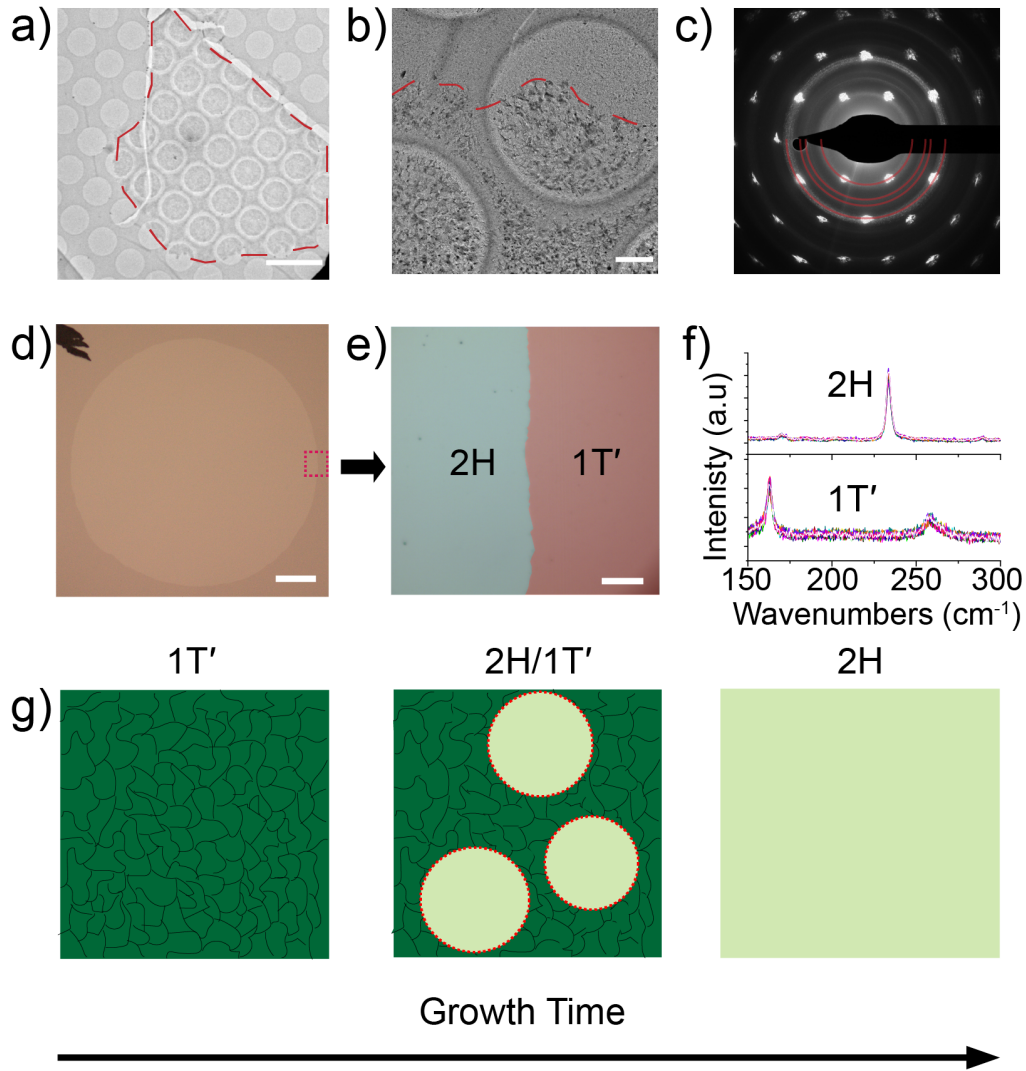

**Figure 5: Phase transition-mediated synthesis of 2H MoTe₂. a)** TEM image shows a large 2H grain in the center of an otherwise 1T′ film. Scale bar, 5 μm. **b)** Higher-magnification TEM image of the boundary shows the border of the 2H grain and the 1T′ region clearly. Scale bar, 500 nm. **c)** Selected area electron diffraction taken from region **(b)** shows one set of diffraction spots for the 2H grain, and diffraction rings for the 1T′ region, suggesting nanoscale grains. The diffraction rings overlaid with the dotted half circles are, from the inner to outer circles, the (200), (210), (310) and (020) planes of the 1T′ MoTe₂ on zone-axis, respectively. **d)** Optical image of 7-8 layer MoTe₂ synthesized at 600 °C for 4 hours, tracking a single 2H grain of size >

1 mm. Scale bar, 200 μm. **e)** Raman map of border of 2H grain and initial 1T' film. Scale bar, 5 μm. **f)** Overlaid spectra from regions shown in **(e)**. 16 spectra shown for each phase. **g)** Nucleation and growth of the 2H phase out of the initial 1T' phase. As long as the Te vapor pressure is maintained, full conversion of 1T' to 2H occurs in the range of 2-4 hours of reaction time at 580 °C and 6-8 hours at 600 °C.

To characterize the film quality, 4-probe devices were fabricated from 3-4 and 7-8 layer thick MoTe₂ films using Cr/Au contacts (Methods). **Figure 6a** shows the resistance of the as-grown 7-8 layer film as temperature was decreased, and Figure 6b shows the log of conductivity as a function of inverse temperature for three samples, which show semiconducting behavior as expected. Using the Arrhenius equation $\sigma = \sigma_0 e^{\frac{-E_A}{K_b T}}$ where $\sigma$ is conductivity, $\sigma_0$ is a temperature independent scaling factor, $E_A$ is the activation energy, and $K_b$ is Boltzmanns constant, an activation energy of 0.14 eV was extracted for the as-grown 7-8 layer thick MoTe₂. This value is much less than the optical bandgap of 1.1eV,[36] but it is similar to the reported activation energies of MoTe₂ flakes exfoliated from bulk,[17] which is attributed to differences in work functions between the contact metal and the film. Thus, transport data suggests that the Fermi energy of our MoTe₂ films converted from MoOₓ films on sapphire must be similar to bulk-exfoliated MoTe₂ flakes. Etching the films with Ar (Methods) resulted in a factor of 4 reduction in the activation barrier: 0.07 eV for the 3-4 layer film and 0.03 eV for the 7-8 layer film. This is attributed to the creation of Te vacancies induced by etching, which have been shown to decrease the Schottky barrier in MoTe₂ flakes exfoliated from bulk.[37] Comparison of Raman spectra before and after Ar etching shows clearly that the MoTe₂ film is degraded after etching (Figure S8, Supporting Information).

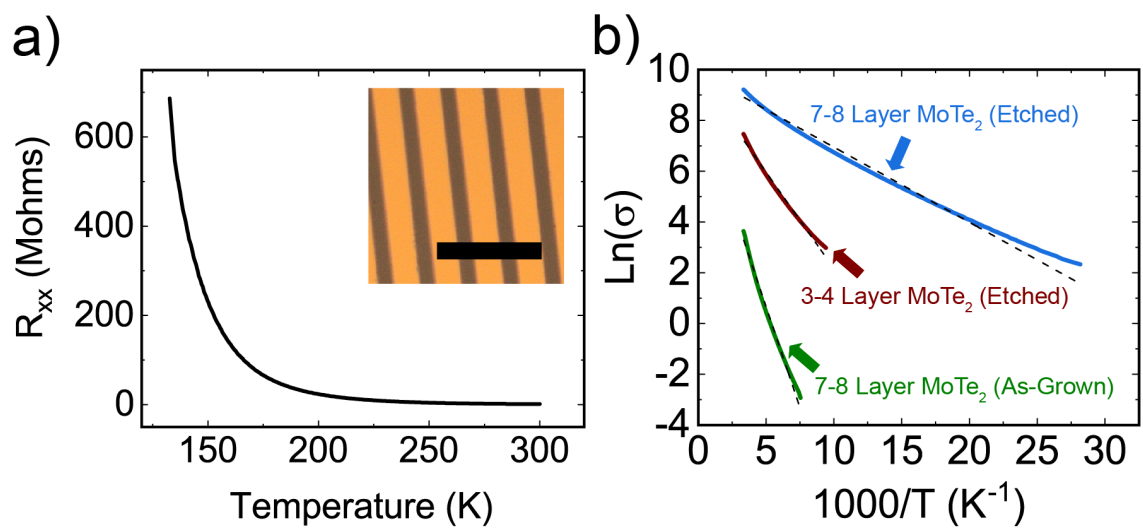

**Figure 6: Temperature-dependent resistance of MoTe$_2$ thin films. a)** Resistance vs. temperature curve for a 7-8 layer thick as-grown MoTe$_2$ device down to 78 K. Inset) Optical image of the device. Scale bar, 200 µm. **b)** Natural log of conductivity ($\ln(\sigma)$) vs. 1000/T, used to extract the activation energy. The green curve denotes data for the 7-8 layer thick as-grown MoTe$_2$ device, and the blue and red curves represent 7-8 and 3-4 layer thick MoTe$_2$ devices after Ar etching. Dotted lines are fits to extract the activation energy.

The choice of sapphire substrate was important to achieve large-area, uniform 2H-MoTe$_2$ thin films with large grains. When the synthesis was repeated using amorphous SiO$_2$ substrates, MoTe$_2$ formed islands instead of continuous films for MoO$_x$ films below 135 cycles (Figure S9, Supporting Information) even though the coverage of MoO$_x$ on sapphire and SiO$_2$ was similar based on LEIS data (Figure S1, Supporting Information). Thus, despite the van der Waals epitaxy which is considered to have weak interactions between the substrate and the 2D film, we note a clear difference in the quality of the resulting MoTe$_2$ film. Our observations are in agreement with previous reports that demonstrate the effect of substrate choice on the subsequent quality of 2D thin films.[38,39] For example, c-plane sapphire or graphene substrates have yielded an epitaxial growth of $WSe_2$ and $NbS_2$ respectively.[40,41] A possible explanation is the difference in adsorption energies of $MoTe_2$ on sapphire and $SiO_2$, which would promote lateral growth of continuous $MoTe_2$ films on sapphire and dewetting of $MoTe_2$ on $SiO_2$. Similar observations have been made for the ALD growth of $WS_2$ and diffusivity of $MoO_3$ on amorphous $Al_2O_3$ versus $SiO_2$ substrates.[42,43]

**Conclusion**

In summary, we have demonstrated a large-area, layer-controlled synthesis of 2H-$MoTe_2$ by tellurizing $MoO_x$ precursor films deposited on sapphire by ALD. We show a monotonic relation between the ALD cycle number of $MoO_x$ and the final thickness of $MoTe_2$, which suggests that $MoTe_2$ films with integer layer thickness can be obtained using $MoO_x$ films of ideal ALD cycle numbers. The synthesis is applicable to other TMDs, including $WTe_2$ (Figure S10, Supporting Information). Our $MoTe_2$ films with layer control enable further opportunities for investigations of layer-dependent properties and phase transitions of $MoTe_2$ and related applications. For example, $MoTe_2$ as phase change memory based on its semiconducting 2H – semimetallic 1T′ phase conversion can be studied as a function of the layer number and substrate effects.

**Methods**

*Atomic Layer Deposition of MoO$_x$ Films*:

Silicon thermal oxide substrates were prepared using a standard RCA process. Samples were immersed in a 5:1:1 volume ratio solution of H$_2$O (18.2 MΩ):NH$_4$OH:H$_2$O$_2$ and subsequently immersed in a 6:1:1 solution of H$_2$O:HCl:H$_2$O$_2$. Both rinses were conducted for 10 min between 70 and 80 °C and were immediately followed by an H$_2$O rinse.

Sapphire substrates were cleaned in a modified version of the RCA process.[44] The first step was a soak in an ethanol bath for 12 hours at room temperature, followed by a rinse with H$_2$O. Samples were then sonicated for 30 min at room temperature in a 1:20:79 solution of detergent:ethanol:H$_2$O. Samples were then soaked in a 3:1 H$_2$SO$_4$:H$_2$O$_2$ piranha solution. Finally, samples were immersed in more concentrated variants of the standard RCA solutions, comprising 2:1:1 H$_2$O:NH$_4$OH:H$_2$O$_2$ and 2:1:1 H$_2$O:HCl:H$_2$O$_2$. Each of the latter 3 rinses was conducted for 20 min at 80 °C and was followed with an H$_2$O rinse. All samples were dried with high purity nitrogen gas between each step.

Atomic layer deposition of MoO$_x$ films was performed in a Cambridge Nanotech S-100 reactor at a substrate temperature of 200 °C. Bis(tert-butylimido)bis(dimethylamido)molybdenum (NtBu)$_2$(NMe$_2$)$_2$Mo and ozone were utilized as precursors in accordance with a previously published procedure.[45] Films of different thicknesses were synthesized using cycle numbers of 5, 14, 27, 60, and 135.

*Conversion of MoO$_x$ to MoTe$_2$ Films*:

MoTe$_2$ thin films were synthesized through the annealing of MoO$_x$ thin films of various thicknesses grown by ALD in a tellurium atmosphere. Te powder (2 g, Sigma-Aldrich, 99.999%) was placed in a 2-inch quartz tube at zone 1 of a two-zone furnace (MTI OTF-1200X-II), while MoO$_x$ thin films on sapphire substrates were placed downstream in the second zone of the tube furnace. After purging the tube many times with house vacuum to ensure no residual oxygen was present, the two zones were heated to 570 °C for Te powder and 580 °C (small 2H grains) or 600 °C (large 2H grains) for MoO$_x$ films in 15 min and held there at times ranging from 1 - 4 hrs depending on the desired ratio of the 2H/1T′ phase. H$_2$ was flowed at 100 sccm at atmospheric pressure during the reaction. After the synthesis was completed, the chamber was purged with 200 sccm Ar gas for 25 minutes, and then rapidly cooled to room temperature by opening the furnace cover.

*Surface Characterization Using LEIS*:

Analysis of the coverage of the MoO$_x$ precursor films was conducted on an ionTOF Qtac high-sensitivity, low-energy ion scattering system. Samples were first subjected to a 15 min cleaning in atomic oxygen *in situ*, and then analyzed using a He$^+$ beam with an energy of 3 keV and a total dose during probing of 1.1x10$^{14}$ ions·cm$^{-2}$. After surface scans, the surface was sputtered with an Ar$^+$ beam with an energy of 500 eV and a dose of 1x10$^{15}$ ions·cm$^{-2}$, which was expected to remove approximately one monolayer.

*Structural and Chemical Characterization*:

Plan-view and cross-section TEM images were taken using a FEI Titan Themis transmission electron microscope with an image corrector at the Advanced Science Research Center at the City University of New York. To avoid sample damage, most TEM images were taken at 80 kV except for Figure 2b, which was taken at 200 kV. For cross-section imaging, a Helios G4 UX DualBeam FIB/SEM from FEI was used to extract lamellas from samples on both $SiO_2$ and sapphire substrates. To protect the samples, a thin (~1 μm) layer of carbon was deposited on the films using an ink pen (Sharpie Brand) and was allowed to dry for ~3 hours under a nitrogen atmosphere. To prevent charging on samples grown on sapphire during FIB milling, a 100 nm layer of gold was deposited using a Leica sputter coater. Inside a FIB, the samples were further coated with 100 nm of electron-beam deposited Pt and 1 μm of ion-beam deposited Pt to protect the area of interest during milling. Energy-dispersive x-ray spectroscopy was taken on the Titan in TEM mode with an energy range of 40 keV. To characterize the surface of the films, atomic force microscopy was conducted using a Cypher ES microscope from Asylum Research. The microscope was operated in peak-force tapping mode. Raman spectroscopy (Horiba) was used to verify film uniformity and characterize layer number for the $MoTe_2$ films. The laser used was 633 nm, with a laser power of 10% and diffraction grating 1800 g/mm. Full conversion of the films from $MoO_x$ to $MoTe_2$ was verified through x-ray photoelectron spectroscopy (XPS) (PHI VersaProbe II).

*Chemical Vapor Transport Synthesis of Bulk Crystal MoTe$_2$*:

Bulk crystal 2H $MoTe_2$ was synthesized in accordance with previously documented methods for the chemical vapor transport (CVT) mediated growth of molybdenum ditelluride[18].

Molybdenum powder (Sigma Aldrich) was mixed in a stoichiometric ratio with tellurium powder (0.5 g/ 1.33 g). That mixture was ground with 2.13 g of NaCl and sealed in a fused silica tube under vacuum in an Ar atmosphere. The tube was placed in a 2-inch quartz tube in a two-zone furnace (MTI OTF-1200X-II) and heated to 1100 °C, and subsequently cooled to 900 °C at a rate of 0.5 °C/hour. The sample was then cooled at a rate of 5 °C/minute until room temperature.

*Device Fabrication and Measurement:*

For as-grown films, 10/100 nm Cr/Au electrodes were thermally deposited on the films using a Kurt Lesker Nano38 thermal evaporator immediately after synthesis. Contact regions were defined using a parallel bar Cu TEM grid as a physical mask, resulting in contact width of 80 μm and channel width of 40 μm. For etched films, $MoTe_2$ films were etched with Ar gas using an Oxford 100 reactive ion etcher for 10 seconds at 50 W. Immediately after etching, 10/100 nm Cr/Au electrodes were deposited using a Denton e-beam evaporator, using the parallel bar Cu TEM grid as the mask. Fabricated devices were measured in both a 4-probe and 2-probe geometry using a Quantum Design Physical Property Measurement System.

**Acknowledgements**
D. H. was supported by the NASA graduate student fellowship #80NSSC19K1131. The fabrication of devices was supported by DOE BES DE-SC0014476. N.C.S and B.D. acknowledge support from the National Science Foundation Grant No. 1605129. The following user facilities are acknowledged for instrument use, scientific and technical assistance: the Yale West Campus Materials Characterization Core (MCC), the Yale Institute for Nanoscience and Quantum Engineering (YINQE), and the Imaging Facility of Advanced Science Research Center at the Graduate Center of CUNY.

**Conflict of Interest**

The authors declare no conflicts of interest.

**Supporting Information Available**: Low-energy ion spectroscopy data showing oxide coverage for MoO$_x$ films, dark-field TEM of individual 2H MoTe$_2$ grains, linear fits of ALD cycle number vs MoTe$_2$ layer number, additional Raman and AFM data to show large-scale uniformity, TEM images of 1T′ MoTe$_2$, Raman spectra showing Ar etching effects, Cross-section TEM comparing MoTe$_2$ growth on SiO$_2$ vs Sapphire, Raman and EDX for converted WTe$_2$ thin film. This material is available free of charge via the Internet at http://pubs.acs.org.